\documentclass[a4paper,12pt]{article}
\usepackage{psfrag}
\usepackage{amssymb}
\usepackage{url}
\usepackage{a4wide,cite,epsfig}
\usepackage{amsmath}
\usepackage{amsfonts}
\usepackage{dcolumn}  
\newcommand{\tpoL}{{2\pi\over L}}
\newcommand{\ovra}[1]{\overset{\rightarrow}{#1}}
\newcommand{\ovla}[1]{\overset{\leftarrow}{#1}}
\newcommand{\ovlra}[1]{\overset{\leftrightarrow}{#1}}

\newcommand{\xiav}{\langle\,\xi\,\rangle}
\parskip\smallskipamount
\def\firstmomentop{O_{\{\rho\mu\}}}
\def\MSbar{{\ensuremath{\overline{\mathrm{MS}}}}}
\def\latt{\mathrm{latt}}
\def\bigG{\frac{g^2 C_\mathrm{F}}{16\pi^2}}
\def\Mmf{M^{\mathrm{MF}}}
\def\wmf{w_0^{\mathrm{MF}}}
\def\Tmf{T_\mathrm{MF}}
\def\gev{\,\mathrm{Ge\kern-0.1em V}}

\begin{document}
\begin{titlepage}
\begin{flushright}
Edinburgh 2006/14\\
SHEP-06-22\\
\end{flushright}
\bigskip
\begin{center}
{\bf {\Large A Lattice Computation of the First Moment of the\\[.3em]
Kaon's Distribution Amplitude}\\[2.5ex]
{\large P.A.~Boyle$^a$, M.A.~Donnellan$^b$, J.M.~Flynn$^b$, A.~J\"uttner$^b$,\\[2mm]
J.~Noaki$^b$, C.T.~Sachrajda$^b$ and R.J.~Tweedie$^a$}\\[3mm]
(UKQCD Collaboration)}\\[2ex]
$^a${\sl Department of Physics and Astronomy, University of
Edinburgh,\\ Edinburgh, EH9 3JZ, UK.}

$^b${\sl
School of Physics and Astronomy, University of Southampton,\\
Southampton, SO17 1BJ, UK.}

\bigskip
\bigskip{\large\bf Abstract}\\[3ex]
\parbox[t]{0.85\textwidth}{
We present a lattice computation of the first moment of the kaon's
leading-twist distribution amplitude. The results were computed
using ensembles with 2+1 dynamical flavours with the domain wall
fermion action and Iwasaki gauge action from the RBC and UKQCD
joint dataset. The first moment is non-zero because of
$SU(3)$-breaking effects, and we find that we are able to measure
these effects very clearly. We observe the expected chiral
behaviour and finally obtain $\xiav(2\,\textrm{GeV})\equiv 3/5
\,a_K^1\,(2\,{\rm GeV})=0.032(3)$, which agrees very well with
results obtained using sum-rules, but with a significantly smaller
error.}
\end{center}
\vspace{1in} PACS: 11.15.Ha, 12.38.Gc, 11.30.Hv, 12.39.St,
14.40.-n
\end{titlepage}

\section{Introduction}
Hadronic light-cone distribution amplitudes are fundamental
non-perturbative ingredients in the QCD analysis of hard exclusive
processes. Phenomenological applications which require knowledge
of the distribution amplitudes include electromagnetic
form-factors at large momentum transfer and related processes
\cite{Chernyak:1977as,Chernyak:1980dj,
Efremov:1979qk,Efremov:1978rn,Chernyak:1977fk,Chernyak:1980dk,Lepage:1980fj}.
More recently, following the development of the factorization
framework, the distribution amplitudes are also an important
component in the phenomenology of exclusive charmless two-body
$B$-decays (i.e. $B$-decays into two light mesons)
~\cite{Beneke:1999br,Beneke:2000ry,Beneke:2001ev,
Bauer:2000ew,Bauer:2000yr,Bauer:2001ct,Bauer:2001yt}. These are a
particularly important set of processes for CKM-analyses and for
studies of CP-violation. Here, we present the first results from
our new lattice project in which we are computing the moments of
the light-cone distribution amplitudes for light pseudoscalar and
vector mesons.

The subject of this letter is the first moment of the
leading-twist distribution amplitude of the kaon, $\phi_K(u,\mu)$,
which pa\-ra\-me\-trizes the overlap of a kaon with longitudinal
momentum $p$ with the lowest Fock state consisting of a quark and
an anti-quark carrying the momentum fractions $up$ and $\bar
up=(1-u)p$, respectively ($u+\bar u=1$). It is defined by the
non-local (light-cone) matrix element:
\begin{equation}\label{eq:phidefz}
 \left.\langle\,0\,|\,\bar{q}(z)\,\gamma_\rho\gamma_5\,{\cal
 P}(z,-z)\,s(-z)\,|\,K(p)\,\rangle
 \right|_{z^2=0}\equiv{f_K}\,(ip_\rho)\,
 \int_0^1du\,e^{i(u-\bar u)p\cdot z}\phi_K(u,\mu)\,,
\end{equation}
where $\mu$ is a renormalization scale and
\begin{equation}\label{eq:pdef} {\cal P}(z,-z)={\cal
 P}\,\exp\left\{-ig\int_{-z}^{z}dw^\mu A_\mu(w)\right\},
\end{equation}
represents the path-ordered exponential from $-z$ to $z$, so that
the bi-local current on the left-hand side of
eq.(\ref{eq:phidefz}) is gauge invariant. The distribution
amplitude is normalized by $\int_0^1du\,\phi_K(u,\mu)=1,$ and can
be expanded in terms of Gegenbauer polynomials $C^{3/2}_n(2u-1)$,
\begin{equation}
 \phi_K(u,\mu)=6u\bar u\left(1+\sum\limits_{n\ge 1}a_n^K(\mu)
    C_n^{3/2}(2u-1)\right).
\end{equation}
The lowest-order anomalous dimensions of the moments $a_n^K(\mu)$
grow with $n$ and thus higher moments may be suppressed for large
values of the renormalization scale $\mu$. In this paper we
present our results for the lowest Gegenbauer moment $a_1^K$,
which is proportional to the average difference of the
longitudinal quark and anti-quark momenta of the lowest Fock
state:
\begin{equation}\label{eq:define_a1K}
a_1^K(\mu)={5\over 3}\int_0^1 du (2u-1)\,\phi_K(u,\mu)
=\frac53\,\langle 2u-1\rangle\equiv\frac53\,\xiav(\mu)\,.
\end{equation}
$a_1^K=5/3\,\xiav$ is obtained from the matrix element of a local
operator,
\begin{equation}\label{eq:1st_m_def_cont}
 \langle\,0\,|\,\bar{q}(0)\,\gamma_\rho\gamma_5
 \,\ovlra{D}_\mu
 \,s(0)\,|\,K(p)\,\rangle = \xiav\,f_K\,p_\rho \,p_\mu=
 {3 \over 5}\,a_1^K f_K\,p_\rho \,p_\mu\,;\\
\end{equation}
eq.(\ref{eq:1st_m_def_cont}) is the leading term in the Taylor
expansion of expression (\ref{eq:phidefz}) around $z=0$. Our
conventions for the covariant derivatives are
$\ovlra{D}_\mu=\ovla{D}_\mu-\ovra{D}_\mu$, $\overset{\rightarrow}
{D}_\mu=\overset{\rightarrow}{\partial}_\mu+ig A_\mu$ and
$\overset{\leftarrow}
{D}_\mu=\overset{\leftarrow}{\partial}_\mu-ig A_\mu$.

The first moment of the kaon's distribution amplitude has in the
past been determined mainly from QCD sum rules, and recent results
include:
\begin{equation}
a_1^K(1\,{\rm GeV})
\,=\,0.05(2)\,\textrm{\cite{Khodjamirian:2004ga},}\quad
0.10(12)\,\textrm{\cite{Braun:2004vf}},\quad 0.050(25)\,
\textrm{\cite{Ball:2005vx}}\quad\textrm{and}\quad
0.06(3)\,\textrm{\cite{Ball:2006fz}}\,.\end{equation}

In this work we present the results of a lattice study of this
quantity using $N_f=2+1$ dynamical flavours of domain wall
fermions~\cite{Kaplan:1992bt,Furman:1994ky}, which have good
chiral properties, and the Iwasaki gauge
action~\cite{Iwasaki:1984cj,Iwasaki:1985we}. We used gauge field
ensembles from the RBC and UKQCD dataset with three values of the
light-quark mass and the calculations are carried out with equal
valence and sea quark masses (i.e. with full unitarity). Further
details of the simulation can be found in sec.\,\ref{sec:Results}
below. We have observed a clear signal for partonic
$SU(3)$-breaking effects in the leading-twist kaon distribution
amplitude and find that the first moment satisfies the chiral
behaviour expected from chiral perturbation theory. For our
\textit{best} results we quote
\begin{eqnarray}\label{eq:best2}
\xiav^{\MSbar}(\mu=2\,\textrm{GeV})&=&0.032\pm 0.003
\qquad(a_1^K(2\,\textrm{GeV})=0.053(5))\\
\xiav^{\MSbar}(\mu=1\,\textrm{GeV})&=&0.040\pm 0.004\, \qquad
(a_1^K(1\,\textrm{GeV})=0.066(6))\,, \label{eq:best1}
\end{eqnarray}
in agreement with most previous results. The errors are already
small, and, as discussed below, will decrease still further in the
near future.

While we were completing this paper, Braun et al. released the
following result from a lattice simulation using improved Wilson
fermions~\cite{Braun:2006dg}:
\begin{equation}\label{eq:braunresult}
a_1^K(2\,\textrm{GeV})=0.0453\pm 0.0009\pm0.0029\,,
\end{equation}
in reasonable agreement with our result.

The plan of the remainder of this paper is as follows. In the
following section we introduce the basic definitions, in
particular the ratio of Euclidean lattice correlation functions
from which we determine $\xiav$ and which we estimate using a
Monte Carlo simulation. Section~\ref{sec:Renormalization} contains
the discussion of the perturbative calculation of the
renormalization constants. We present our results for the bare
matrix elements in section~\ref{sec:Results} and combine them with
the renormalization constants and discuss the systematic
uncertainties in section~\ref{sec:final}, where we present our
final result. We end with a brief summary and conclusions
(sec.~\ref{sec:concs}).

\section{\boldmath{$\xiav^{\bf bare}$ from Lattice Correlation Functions}}
\label{sec:Lattice study}
We start by briefly describing the overall strategy of our
calculation of $\xiav$. As we explain in this section, we exploit
the fact that $\xiav$ can be obtained directly from a ratio of two
Euclidean correlation functions, which we evaluate using a Monte
Carlo simulation. The statistical fluctuations are reduced in the
ratio and, for each choice of quark masses used in the simulation,
we are able to obtain $\xiav$ with good precision.

In constructing the lattice operators, we use the following
symmetric left- and right-acting covariant derivatives:
\begin{equation}
 \overset{\rightarrow}{D}_\mu\psi(x)=\frac{1}{2a}\left\{\,U(x,x+\hat\mu)
    \psi(x+\hat\mu)- U(x,x-\hat\mu)\psi(x-\hat\mu)\,\right\}\,,
\end{equation}
and
\begin{equation}
 \bar\psi(x)\overset{\leftarrow}{D}_\mu=\frac{1}{2a}\left\{\bar{
 \psi}(x+\hat\mu)U(x+\hat\mu,x)- \bar{
 \psi}(x-\hat\mu)U(x-\hat\mu,x)\,\right\}\,.
\end{equation}
where the $U$'s are the gauge links and $\hat\mu$ is a vector of
length $a$ in the direction $\mu$ ($a$ denotes the lattice
spacing).

To illustrate the method, consider the local lattice
operators\,\footnote{In the simulation we actually use a smeared
pseudoscalar density $P$ in order to improve the overlap with the
kaon state. The present discussion holds for both local and
smeared pseudoscalar densities.}:
\begin{equation}\label{eq:operatorsdef}
 O_{\rho\mu}(x)=\bar{q}(x)\gamma_\rho\gamma_5\,\ovlra{D}_\mu\,
    s(x)\,,\quad
 A_{\rho}(x)=\bar{q}(x)\,\gamma_\rho\gamma_5\,
 s(x)\quad\textrm{and}\quad
 P(x)=\bar{q}(x)\,\gamma_5\, s(x)\,,
\end{equation}
and define the two-point correlation functions
\begin{equation}\label{eq:define_C_rhomu}
 \begin{array}{rcl}
  C_{\rho\mu}(t,{\vec p}\,)=\sum\limits_{\vec x}e^{i \vec{p}\cdot\vec{x}}
  \langle 0|O_{\rho\mu}(t,\vec{x}\,) P^\dagger(0)|0\rangle\,,
 \end{array}
\end{equation}
and
\begin{equation}\label{eq:define_C_AP}
 \begin{array}{rcl}
  C_{A_\nu P}(t,\vec{p}\,)=\sum\limits_{\vec x}e^{i\vec{p}\cdot\vec{x}}
  \langle 0|A_\nu(t,\vec{x}\,) P^\dagger(0)|0\rangle\,.
 \end{array}
\end{equation}
Here $q$ and $s$ represent the light and strange quark fields,
respectively. At large Euclidean times $t$ and $T-t$, where $T$ is
the length of the lattice in the time direction, the correlation
functions (\ref{eq:define_C_rhomu}) and (\ref{eq:define_C_AP})
approach
\begin{eqnarray}\label{cfs}
C_{\{\rho\mu\}}(t,\vec{p}\,)&\to&
  \frac{Z_P\,f_K^{\rm bare}\,
  e^{- E_{K}\, T/2} \,\sinh((t-T/2)E_{K})}{E_K}\times
  (ip_\rho) (ip_\mu)\,\xiav^{\rm
  bare}\,,
  \\
  C_{A_\nu P}(t,\vec{p}\,)&\to&
  \frac{Z_P\,f_K^{\rm bare}\,
  e^{- E_{K}\, T/2} \,\sinh((t-T/2)E_{K})}{E_K}
\times (ip_\nu)
\end{eqnarray}
where we have used (\ref{eq:define_a1K}),
(\ref{eq:1st_m_def_cont}) and defined $Z_P\equiv\langle
K(p)|P^\dagger|0 \rangle$ and the kaon's bare decay constant
$\langle0|A_\nu|K(p)\rangle\equiv i p_\nu f_K^{\rm bare}$. The
superscript \textit{bare} denotes the fact that the operators are
the bare ones in the lattice theory with ultraviolet cut-off
$a^{-1}$. Taking the ratio of the two correlation functions
\begin{equation}\label{eq:ratio_R}
 R_{\{\rho\mu\};\,\nu}(t,\vec{p}\,)\,\equiv\,\frac
 {C_{\{\rho\mu\}}(t,\vec{p}\,)}
 {C_{A_\nu P}(t,\vec{p}\,)}
\,\to\,
 \,i\,{p_\rho p_\mu \over p_\nu}\,\xiav^\textrm{bare},
\end{equation}
allows us readily to extract the bare value of $\xiav$. The braces
in the subscripts $\{\rho\mu\}$ indicate that the indices are
symmetrized.

In the continuum $\ovlra{\cal O}_{\{\rho\mu\}}$ transforms as a
second rank Lorentz tensor, whereas on the lattice we need to
consider transformation properties under the hypercubic group and
discrete symmetries, in particular parity and charge conjugation.
We choose to evaluate the matrix elements of  $\ovlra{\cal
O}_{\{\rho\mu\}}$ for $\mu\neq\rho$. These operators transform as
a six-dimensional representation of the hypercubic group and
lattice symmetries exclude mixing with other
operators\cite{Gockeler:1996mu} including operators containing a
total derivative (since we are considering matrix elements with a
non-zero momentum transfer, operators with a total derivative have
to be considered). As can be seen from eq.\,(\ref{eq:ratio_R}), in
order for the matrix elements to be non-zero, we require both
$p_\mu\neq0$ and $p_\rho\neq0$. We satisfy this condition by
taking $\mu=\nu=4$, $\rho=1,2$ or 3 and $|p_\rho|=2\pi/L$\,.

Having obtained $\xiav^\textrm{bare}$, we need to determine the
renormalization factor relating the lattice operators $\ovlra{\cal
O}_{\{\rho\mu\}}$ and $A_{\nu}$ to the corresponding continuum
operators in some standard renormalization scheme; here we do this
at one-loop order in perturbation theory as explained in the
following section.

\section{Perturbative Renormalization of the Lattice Operators}
\label{sec:Renormalization}

\begin{figure}
 \begin{center}
  \epsfig{scale=0.6,file=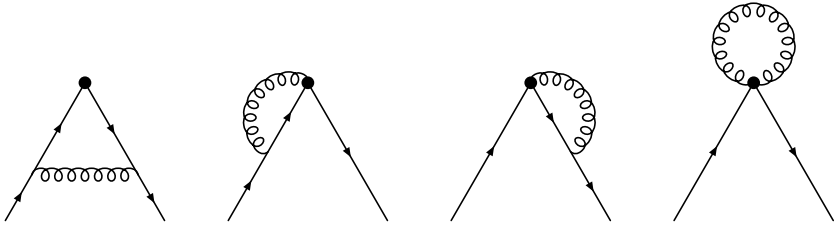}
  \caption{One-loop vertex diagrams evaluated in the perturbative renormalization
   of $O_{\{\rho\mu\}}$}\label{Fig:vertex}
  \label{RGIfac}
 \end{center}
\end{figure}

\begin{figure}
 \begin{center}
  \epsfig{scale=0.6,file=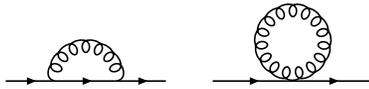}
  \caption{One-loop diagrams contributing to the quarks' wave function renormalization.}
  \label{Fig:wf}
  \end{center}
\end{figure}

The perturbative matching from the lattice to $\MSbar$ schemes is
performed by comparing one-loop calculations of the amputated
two-point Green function with an insertion of the operator
$\firstmomentop$ in both schemes (which requires the evaluation of
the diagrams in fig.\,\ref{Fig:vertex}), together with appropriate
wave function renormalization factors (fig.\,\ref{Fig:wf}).
Defining $\firstmomentop^\MSbar(\mu) = Z_{\firstmomentop}
\firstmomentop^\latt(a)$, the renormalization factor is given by
\begin{equation}
\label{eq:Zpt} Z_{\firstmomentop} = \frac1{(1-w_0^2)Z_w}
 \left[ 1 + \bigG \left( -\frac83 \ln(\mu^2 a^2) + \Sigma_1^\MSbar
        - \Sigma_1 + V^\MSbar - V \right) \right].
\end{equation}
In this expression, $(1-w_0^2)Z_w$ is a characteristic
normalization factor for the physical quark fields in the domain
wall formalism. It is a common factor in the numerator and
denominator of the ratio $R_{\{\rho\mu\};\nu}$ as are the
contributions from the wave function renormalization. $Z_w$
represents an additive renormalization of the large Dirac mass or
domain wall height $M=1-w_0$ which can be rewritten in
multiplicative form at one-loop as
\begin{equation}
Z_w = 1+\bigG z_w, \qquad z_w = \frac{2w_0}{1-w_0^2}\,\Sigma_w.
\end{equation}
The one-loop correction $z_w$ becomes very large for certain
choices of $M$~\cite{Aoki:1998vv,Aoki:2002iq}, including that used
in our numerical simulations, so that some form of mean-field
improvement is necessary, as discussed below~\footnote{The factor
$1/(1-w_0^2)Z_w$ cancels however, in the evaluation of the ratio
$Z_{\firstmomentop}/Z_A$, where $Z_A$ is the renormalization
constant for the axial current.}.

The terms $\Sigma_1^\MSbar$ and $\Sigma_1$ come from quark wave
function renormalization. The terms $V^\MSbar$ and $V$ come from
the one-loop corrections to the amputated two-point function. They
are given by ``vertex'' and ``sail'' diagrams, plus an operator
tadpole diagram in the lattice case. Using naive dimensional
regularization (NDR) in Feynman gauge with a gluon mass infrared
(IR) regulator,
\begin{equation}
\Sigma_1^\MSbar = \frac12, \qquad V^\MSbar = -\frac{25}{18}.
\end{equation}
The lattice contribution $\Sigma_1$ has been evaluated for domain
wall fermions with the Iwasaki gluon action in Feynman gauge and a
gluon mass IR regulator in~\cite{Aoki:2002iq}. We have calculated
the lattice vertex term $V$ for the same action, gauge and IR
regulator to complete the evaluation of $Z_{\firstmomentop}$. The
perturbative calculation is explained
in~\cite{Aoki:1998vv,Aoki:2002iq,Capitani:2005vb} and the form of
the Iwasaki gluon propagator can be found
in~\cite{Iwasaki:1983ck}. Values for $V$ are given as a function
of $M$ in Table~\ref{tab:VvsM}. Chiral symmetry of the domain wall
action implies that these results also apply for the operator
which is $\firstmomentop$ without the $\gamma_5$. We have
confirmed that our results reproduce those found by
Capitani~\cite{Capitani:2005vb} if we replace the gluon propagator
for the Iwasaki gauge action by the propagator corresponding to
the standard plaquette action. This provides a powerful check of
our calculation. We note that the perturbative renormalization
factor for the same operator using overlap fermions and the
L\"uscher--Weisz gauge action can be found
in~\cite{Horsley:2005jk}.
\begin{table}[t]
\def\arraycolsep{0.75em}
\[
\begin{array}{D..1D..4D..4D..4D..4D..4}
 \multicolumn1cM &
 \multicolumn1c{z_w} &
 \multicolumn1c{z_w^\mathrm{MF}} &
 \multicolumn1c{\Sigma_1} &
 \multicolumn1cV &
 \multicolumn1c{\delta\Sigma_1+\delta V} \\
\hline
 0.1 & -243.86 & -86.579 & 4.6519 & -4.6297 & -0.9110 \\
 0.2 & -113.29 & -39.501 & 4.5193 & -4.5614 & -0.8468 \\
 0.3 & -69.404 & -23.830 & 4.4093 & -4.5101 & -0.7881 \\
 0.4 & -47.077 & -15.949 & 4.3158 & -4.4678 & -0.7369 \\
 0.5 & -33.278 & -11.142 & 4.2354 & -4.4311 & -0.6932 \\
 0.6 & -23.648 & -7.8365 & 4.1665 & -4.3980 & -0.6574 \\
 0.7 & -16.300 & -5.3538 & 4.1079 & -4.3673 & -0.6295 \\
 0.8 & -10.263 & -3.3459 & 4.0593 & -4.3381 & -0.6101 \\
 0.9 & -4.9617 & -1.6078 & 4.0204 & -4.3097 & -0.5996 \\
 1.0 & 0.0 & 0.0 & 3.9915 & -4.2816 & -0.5988 \\
 1.1 & 4.9442 & 1.5902 & 3.9731 & -4.2529 & -0.6090 \\
 1.2 & 10.192 & 3.2748 & 3.9664 & -4.2232 & -0.6321 \\
 1.3 & 16.136 & 5.1900 & 3.9727 & -4.1916 & -0.6700 \\
 1.4 & 23.346 & 7.5350 & 3.9943 & -4.1571 & -0.7261 \\
 1.5 & 32.784 & 10.648 & 4.0343 & -4.1182 & -0.8050 \\
 1.6 & 46.322 & 15.194 & 4.0974 & -4.0728 & -0.9135 \\
 1.7 & 68.294 & 22.720 & 4.1905 & -4.0176 & -1.0618 \\
 1.8 & 111.69 & 37.901 & 4.3249 & -3.9462 & -1.2676 \\
 1.9 & 241.55 & 84.270 & 4.5209 & -3.8447 & -1.5651
\end{array}
\]
\caption{Constants needed for the perturbative renormalization of
the
  operator $\firstmomentop$ using domain wall fermions and the Iwasaki
  gauge action ($c_1=-0.331$). $M$ is the domain wall height and
  $\delta\Sigma_1+\delta V = \Sigma_1^\MSbar - \Sigma_1 + V^\MSbar -
  V$, while other quantities are defined in the text. $\Sigma_1$ and
  $V$ are dependent on the gauge and the infrared regulator: Feynman
  gauge and a gluon mass are used here. $z_w^{(\mathrm{MF})}$ and
  $\Sigma_1$ are extracted from the results in~\cite{Aoki:2002iq},
  while $V$ has been calculated as part of this work.}
\label{tab:VvsM}
\end{table}

Our numerical simulations use $M=1.8$. For this value of $M$, with
the Iwasaki gluon action, the one-loop coefficient $z_w$ in the
physical quark normalization can be extracted from $\Sigma_w$ in
Table~III of~\cite{Aoki:2002iq}. This gives $z_w \approx 112$,
making it clear that mean-field improvement is necessary. We
follow the prescription described in~\cite{Aoki:2002iq}.

The first step is to define a mean-field value for the domain wall
height,
\begin{equation}
\Mmf = M - 4(1-P^{1/4})
\end{equation}
where $P=0.58813(4)$ is the average plaquette value in our
simulations. This leads to
\begin{equation}
\Mmf = 1.3029.
\end{equation}
The physical quark normalization factor becomes
$\left[1-(\wmf)^2\right]Z_w^\mathrm{MF}$, with
\begin{equation}
Z_w^\mathrm{MF} = 1+\bigG z_w^\mathrm{MF}, \qquad z_w^\mathrm{MF}
= \frac{2\wmf}{1-(\wmf)^2}\,(\Sigma_w + 32\pi^2\Tmf)
                = 5.2509,
\end{equation}
where $\Tmf=0.0525664$~\cite{Aoki:2002iq} is a mean-field tadpole
factor and $\Sigma_w$ is evaluated at $\Mmf$, leading to
$z_w^\mathrm{MF}=5.2509$. Likewise, $\Sigma_1=3.9731$ and
$V=-4.1907$ in equation~(\ref{eq:Zpt}) are evaluated at $\Mmf$ and
the mean-field improved renormalization factor for our simulations
becomes:
\begin{equation}
Z_{\firstmomentop} = \frac1{0.9082}\,
 \left[1-\bigG \, 5.2509\right]
 \left[1+\bigG \left( -\frac83 \ln(\mu^2 a^2) -0.6713\right) \right].
\end{equation}
\begin{center}
\end{center}

We make two choices for the mean-field improved $\MSbar$ coupling.
The first uses the measured plaquette value, $P$, according
to~\cite{Aoki:2002iq}
\begin{equation}
\frac1{g^2_\MSbar(\mu)} =
 \frac P{g^2} + d_g + c_p + \frac{22}{16\pi^2}\,\ln(\mu a)\,,
\end{equation}
where $d_g=0.1053$ and $c_p=0.1401$ for the Iwasaki gauge action
and $\beta = 6/g^2 = 2.13$ in our simulations. The second choice
is the usual continuum $\MSbar$ coupling. At $\mu a = 1$, we find
$\alpha_\MSbar(\mathrm{plaq}) = 0.1752$ and
$\alpha_\MSbar(\mathrm{ctm}) = 0.3385$. This disparity in the
values of the two couplings is further motivation for the
programme of non-perturbative renormalization which we are
currently undertaking. With these two choices of coupling, our
value for the renormalization factor becomes:
\begin{equation}
Z_{\firstmomentop} = \begin{cases}
 0.9811 & \mbox{plaquette coupling}\\
 0.8719 & \mbox{continuum \MSbar}\ .
             \end{cases}
\end{equation}

We also evaluate the mean-field improved expression for the axial
vector current~\cite{Aoki:2002iq}, interpolating to our mean-field
$\Mmf$, and obtain
\begin{equation}
Z_\mathrm{A} =  \begin{cases}
 0.7947 & \mbox{plaquette coupling}\\
 0.6514 & \mbox{continuum \MSbar}\ .
             \end{cases}
\end{equation}
The ratio of the two renormalization factors is
\begin{equation}\label{eq:zrpert}
\frac{Z_{\firstmomentop}}{Z_\mathrm{A}} =  \begin{cases}
 1.2346 & \mbox{plaquette coupling}\\
 1.3384 & \mbox{continuum \MSbar}\ .
             \end{cases}
\end{equation}
For the purposes of this letter we include the spread of results
in eq.(\ref{eq:zrpert}) as the estimate of our current systematic
uncertainty of the renormalization factor~\footnote{We mention in
passing that using the bare coupling,
$\frac{Z_{\firstmomentop}}{Z_\mathrm{A}}=1.26$\,.}. This
uncertainty will be significantly reduced as we complete our
programme of non-perturbative renormalization.
$\xiav^\textrm{bare}$ should be multiplied by the factor on the
right-hand side of eq.(\ref{eq:zrpert}) to obtain the result in
the $\overline{\textrm{MS}}$ scheme. For this factor we take
\begin{equation}\label{eq:zav}
\frac{Z_{\firstmomentop}}{Z_\mathrm{A}}=1.28\pm 0.05\,.
\end{equation}

\section{Numerical Simulation and Results}\label{sec:Results}
The numerical results presented here are based on Monte Carlo
estimates of correlation functions evaluated on representative
sets of UKQCD/RBC gauge field configurations that were generated
with $N_f=2+1$ flavours of dynamical domain wall fermions
\cite{Kaplan:1992bt,Furman:1994ky} with Iwasaki Gauge action
\cite{Iwasaki:1984cj,Iwasaki:1985we} using the QCDOC computer
\cite{qcdoc1,qcdoc2,qcdoc3,qcdoc4}. The hadronic spectrum and
other properties of these configurations have been studied in
detail and the results will be presented in
ref.\,\cite{configpaper}. We wish to thank our collaborators in
the RBC and UKQCD collaborations for access to several preliminary
results which we require for this study. These are specifically
$m_{res}$ in the chiral limit, the bare kaon pseudoscalar masses,
and the inverse lattice spacing.

The preliminary nature of these intermediate values does not
introduce a significant uncertainty in our results compared to our
statistical errors and the systematic errors due to the chiral
extrapolation and renormalization. Indeed we shall only quote the
lattice spacing to two significant figures and demonstrate the
insensitivity of our results to this. However, we caution the
reader that definitive results for these intermediate quantities
will be found in the forthcoming publication.

The lattice volume is $(L/a)^3\times T/a=16^3\times32$ and the
length of the fifth dimension is $L_s=16$. The choice of bare
parameters is $\beta=2.13$ for the bare gauge coupling,
$am_s=0.04$ for the strange quark mass (which has been tuned to
correspond to the physical value) and $am_q=0.03,\,0.02,\,0.01$
for the bare light-quark masses. With this choice of simulation
parameters the lattice spacing is
$a^{-1}=1.6$\,GeV~\cite{configpaper}. Due to the remnant chiral
symmetry breaking the quark mass has to be corrected additively by
the residual mass in the chiral limit, $am_{\rm
res}=0.003$~\cite{configpaper}.

Statistical errors for observables have been estimated both with
the jack-knife and with a direct estimation of the integrated
autocorrelation time as suggested in \cite{Wolff:2003sm}.

\subsection{Bare correlation functions}
For each value of the light-quark mass we computed the correlation
functions on 300 gauge configurations separated by 10 trajectories
in the Monte Carlo history. On each configuration we averaged the
results obtained from 4 sources for the lightest quark mass
($m_qa=0.01$) and 2 sources for the remaining two masses
($m_qa=0.02$ and 0.03). The sources were chosen to be at the
origin and (8,8,8,16) for all three masses, and in addition at
(4,4,4,8) and (12,12,12,24) for $m_qa=0.01$. In order to improve
the overlap with the ground state, at the source where we insert
the density $P^\dagger$, we employed gauge invariant Jacobi
smearing \cite{Allton:1993wc} (radius 4 and 40 iterations) with
APE-smeared links in the covariant Laplacian operator ($4$ steps
and smearing factor $2$) \cite{Falcioni:1984ei,Albanese:1987ds}.

The preliminary kaon masses corresponding to the simulated bare
light-quark masses are $am_K^{0.03}=0.4164(10)$,
$am_K^{0.02}=0.3854(10)$, and
$am_K^{0.01}=0.3549(14)$~\cite{configpaper}.

\begin{figure}
 \begin{center}
  \psfrag{tovera}[t][c][1][0]{$t/a$}
  \psfrag{1stmoment}[c][t][1][0]{$\xiav^{\textrm{bare}}$}
  \psfrag{Legendmass1}[c][c][1][0]{$am_{ud}=0.01$}
  \psfrag{Legendmass2}[c][c][1][0]{$am_{ud}=0.02$}
  \psfrag{Legendmass3}[c][c][1][0]{$am_{ud}=0.03$}
  \epsfig{scale=.4,angle=270,file=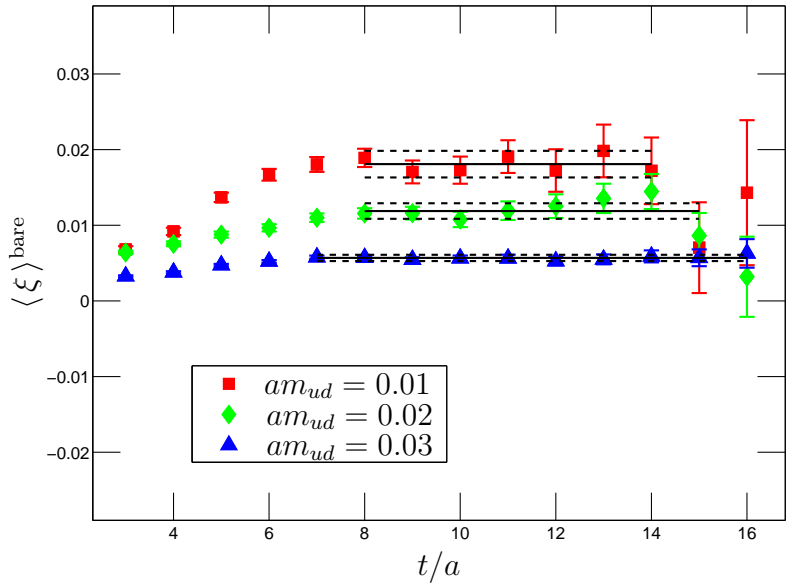}\\
 \end{center}
 \caption{Jack-knife results for $\xiav^{\textrm{bare}}$ as a function of the time.
The ranges over which we fit and the corresponding results are
indicated by the black lines.}\label{fig:1st_cf}
\end{figure}

In order to extract $\xiav$ from the ratio
$R_{\{\rho\mu\};\,\nu}(t,{\vec p}\,)$ defined in
(\ref{eq:ratio_R}) we need $|{\vec p}\,|\neq 0$. Since hadronic
observables with larger lattice momenta have larger lattice
artefacts and statistical errors, we restrict the choice of
indices to $\rho=\nu=4$~\footnote{The index $4$ corresponds to the
time-direction.} and $\mu=1,2,3$ with $|{\vec p}\,|=2\pi/L$
($p_\mu=\pm 2\pi/L$ with the remaining two components of ${\vec
p}$ equal to 0). $\xiav^{\textrm{bare}}$ is then obtained from the
correlation function at large times:
\begin{equation}
R_{\{4 k\};\,4}(t,p_k=\pm 2\pi/L)= \pm\,
i\,\tpoL\,\xiav^{\textrm{bare}}\,,  \quad
|\vec{p}\,|=\tpoL\,,\quad k=1,2,3\,.
\end{equation}
The plot in figure \ref{fig:1st_cf} shows our results for $\xiav$
as a function of $t$ obtained from  the ratio $R_{\{4
k\};\,4}(t,p_k=\pm 2\pi/L)$ for the three values of the mass of
the light quark. The results have been averaged over the three
values for $k$ and the 6 equivalent lattice momenta with $|\vec
p\,|=2\pi/L$ and combining the results at $t$ with those at
$T-t-1$. There are clear plateaus, demonstrating that the
$SU(3)$-breaking effects are measurable and $\xiav$ can be
determined. The results from the fits for $\xiav^{\textrm{bare}}$
are summarized in table \ref{tab:results_first_moment}.
\begin{table}
 \begin{center}
 \begin{tabular}{cccc||c}
  $am_{ud}$&$0.03$  &$0.02$     &$0.01$&$\chi$-limit\\[1mm]
  \hline\hline&&&&\\[-3mm]
  $\langle u-\bar u\rangle$&
                0.0057(4) &0.0119(10)&0.0181(18)    &0.0262(23)\\
 \end{tabular}
 \caption{Summary of results for the bare values of the 1st moment
 of the kaon's distribution amplitude. The result we obtain after the linear chiral
 extrapolation is
quoted in the right-most
 column. }\label{tab:results_first_moment}
 \end{center}
\end{table}

\subsection{Chiral extrapolation}\label{subsec:chiral}
For the pion the first moment $a_1^\pi$ vanishes since isospin
symmetry induces invariance under $u\leftrightarrow(1-u)$. For a
non-degenerate quark-antiquark pair (such as the kaon) flavour
symmetry breaking implies that the first moment of the
distribution amplitude is non-zero. The leading $SU(3)$-violating
effects for the kaon's distribution amplitude have been studied at
next-to-leading order in chiral perturbation theory ($\chi$PT) in
\cite{Chen:2003fp,Chen:2005js}. No chiral logarithms appear and the prediction
for the mass-dependence is
\begin{equation}\label{eq:ChPT}
 \xiav = \frac{8B_0}{f^2}(m_s-m_{ud})b_{1,2}\,,
\end{equation}
where $f$ and $B_0$ are conventional $\chi$PT parameters and
$b_{1,2}$ is a Wilson coefficient as introduced in
\cite{Chen:2003fp}.

\begin{figure}
 \begin{center}
  \psfrag{mqplusmres}[t][c][1][0]{$am_q+am_{\rm res}$}
  \psfrag{fstmom}[c][t][1][0]{$\xiav^{\textrm{bare}}$}
  \psfrag{Legendmass1}[c][c][1][0]{$am_{ud}=0.01$}
  \psfrag{Legendmass2}[c][c][1][0]{$am_{ud}=0.02$}
  \psfrag{Legendmass3}[c][c][1][0]{$am_{ud}=0.03$}
  \epsfig{scale=.4,angle=270,file=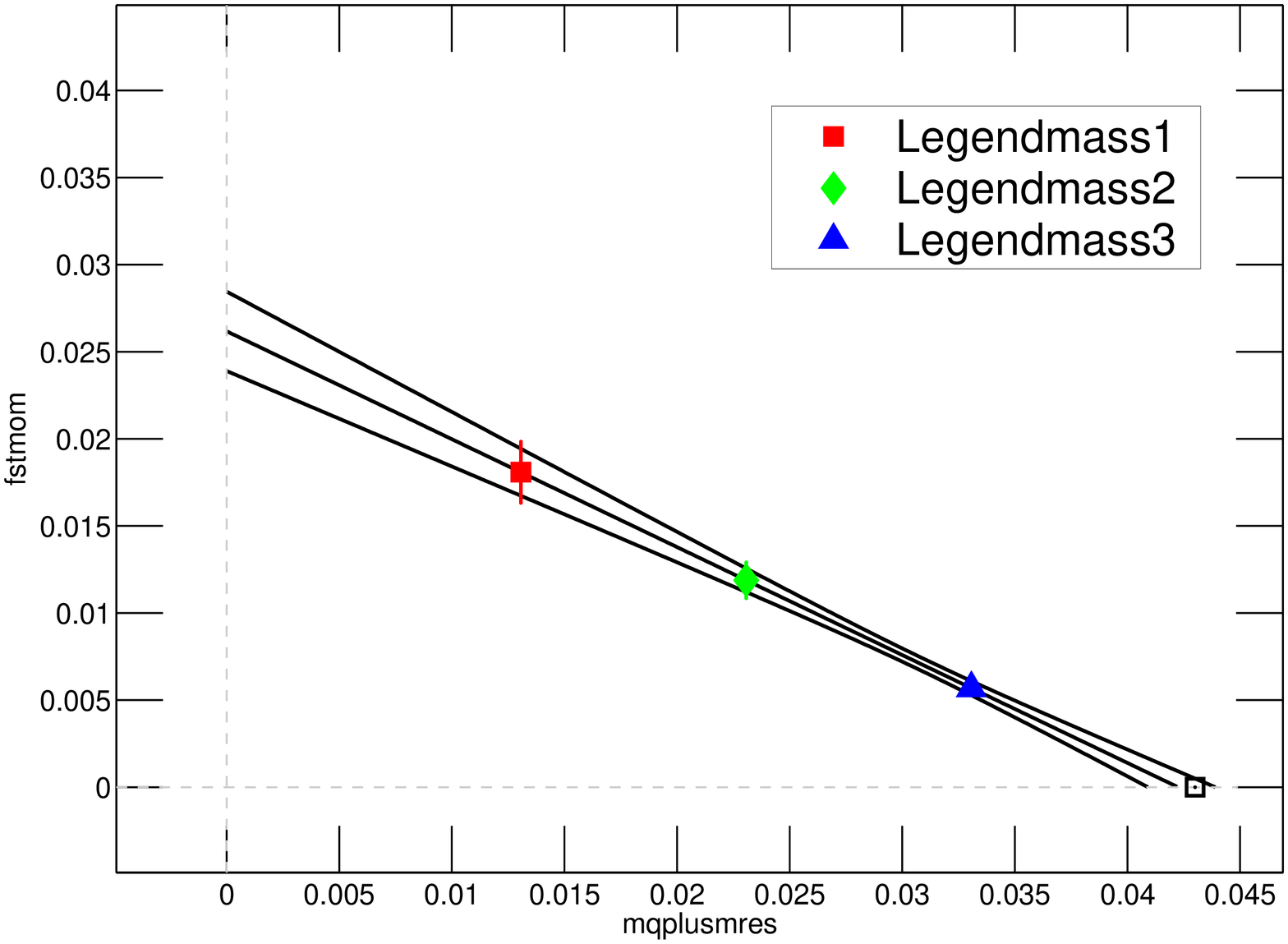}\\
 \end{center}
 \caption{Linear chiral extrapolation for $\xiav^{\textrm{bare}}$.}
\label{fig:1st_ch_extrapol}
\end{figure}

We plot our results for $\xiav^\textrm{bare}$ as a function of the
light-quark mass in fig.~\ref{fig:1st_ch_extrapol}. We take into
account the remnant chiral symmetry breaking by defining the
chiral limit at the point $am_q+am_{\rm res}=0$. The linear
behaviour predicted in eq.\,(\ref{eq:ChPT}) is well satisfied
(with a tiny $\chi^2/$d.o.f. of about $10^{-5}$). Moreover the
line passes through $\xiav^\textrm{bare}=0$ at a value of the
light-quark mass (denoted by the open square in
fig.~\ref{fig:1st_ch_extrapol}) which is consistent with the mass
of the strange quark, as expected for the $SU(3)$ symmetric case
($am_{ud}=am_{s}=0.04$). More specifically, the intercept of the
linear fit with the $\xiav^\textrm{bare}=0$ axis occurs at
$am_{ud}=0.0391^{+0.0017}_{-0.0013}$\,.

From the linear fit~\footnote{We have also performed quadratic
fits to the chiral behaviour but the results do not change in any
significant way.} we obtain $\xiav^\textrm{bare}=0.0262(23)$ in
the chiral limit and in the next section we combine this result
with the renormalization constant in eq.(\ref{eq:zav}) to arrive
at our final result.

\section{Systematic Uncertainties and our Final
Result}\label{sec:final} To obtain the final result for the
$\xiav$ in the $\MSbar$ scheme at $\mu\simeq 1.6$\,GeV we multiply
the result obtained from the bare operators with cut-off
$a^{-1}=1.6$\,GeV, $\xiav^{\textrm{bare}}=0.0262(23)$, by the
ratio of renormalization factors in eq.(\ref{eq:zav}),
$\frac{Z_{\firstmomentop}}{Z_\mathrm{A}}=1.28(5)$:
\begin{equation}\label{eq:resulta}
\xiav^{\MSbar}(\mu=1.6\,\textrm{GeV})=0.034\pm
0.003\,.\end{equation}

In order to be able to compare our result with previous
calculations we evolve it to renormalization scales of 1\,GeV and
2\,GeV using the three-loop anomalous
dimension~\cite{Larin:1993vu}, obtaining:
\begin{eqnarray}\label{eq:final2}
\xiav^{\MSbar}(\mu=2\,\textrm{GeV})&=&0.032\pm 0.003\\
\xiav^{\MSbar}(\mu=1\,\textrm{GeV})&=&0.040\pm 0.004\,.
\label{eq:final1}
\end{eqnarray}
The error in the renormalization factor due to the uncertainty in
the lattice spacing is negligible. For example if we
conservatively allow the lattice spacing to vary between 1.58\,GeV
and 1.62\,GeV, the contribution to the relative error on
$\xiav^{\MSbar}$ is less than 0.2\%.

Among the uncertainties which, at this stage at least, we are not
in a position to check numerically are the continuum
extrapolation, finite-volume effects and the fact that the strange
quark mass ($m_sa=0.04$) is only approximately tuned to its
physical value. The lattice artefacts are formally of
$O(a^2\Lambda^2_{\textrm QCD})\simeq 2.5\%$. We would expect the
finite volume effects to be small and are currently checking this
with a simulation on a $24^3\times 64$ lattice. The strange quark
mass appears to be well tuned~\cite{configpaper} so again we
expect the contribution to the error from this uncertainty to be
very small. Thus we expect the errors from these three sources to
be sufficiently small not to change the errors quoted in
eqs.(\ref{eq:final2}) and (\ref{eq:final1}) which we take to be
our best estimates. We are also carrying out a systematic
programme of non-perturbative renormalization which will enable us
to reduce the uncertainty in the renormalization constants.

\section{Summary and Conclusions}\label{sec:concs}
In this letter we have presented the first results from our major
lattice study of the leading-twist light-cone distribution
amplitudes of the light mesons. We demonstrate that the
$SU(3)$-breaking effects which lead to a non-zero value for the
first moment of the kaon's distribution amplitude are sufficiently
large to be calculable in lattice computations and satisfy the
expected chiral behaviour. Our results for the first moment are
presented in eqs.(\ref{eq:final2}) and (\ref{eq:final1}).

We are in the process of implementing a number of improvements,
including non-pert\-urb\-ative renormalization and a simulation on
a larger lattice ($24^3\times 64$) which will help reduce these
two sources of systematic error (or in the latter case give us
confidence in our expectation that the finite-volume errors are
indeed small). We will also produce results for the second moment
of the pion's and kaon's distribution amplitudes and for those of
the $\rho$ and $K^\ast$ vector mesons. In order to reduce the
lattice artefacts we will investigate the use of partially twisted
boundary conditions~\cite{sv,bc,fjs} which will allow us to
calculate the observables at smaller values of lattice momenta.

While we were completing this paper, ref.\,\cite{Braun:2006dg}
appeared with the results of an $N_f=2$ study of the distribution
amplitude, using $O(a)$ improved Wilson fermions and the plaquette
Wilson gauge action. For the first moment of the kaon's
distribution amplitude the calculation was performed at
$\beta=5.29$ corresponding to a lattice spacing of about 2.6\,GeV.
The result of ref.\,\cite{Braun:2006dg} is given in
eq.\,(\ref{eq:braunresult}) above.

\section*{Acknowledgements} All gauge configurations were generated on QCDOC using the
Columbia Physics System \cite{CPS}, and correlation functions made use of
the CHROMA QCD library and SciDAC software stack \cite{Edwards:2004sx}.
Both code bases use BAGEL high performance code \cite{Bagle}.

We are grateful to all our colleagues in the RBC and UKQCD
collaborations who have contributed to the research programme in
which these simulations were performed. These include David
Antonio, Norman Christ, Mike Clark, Paul Cooney, Chulwoo Jung,
Richard Kenway, Shu Li, Meifeng Lin, Bob Mawhinney, Chris Maynard,
Brian Pendleton, Azusa Yamaguchi and James Zanotti. Finally we
wish to thank the RBC and UKQCD collaborations for assistance in
expediting this paper via preliminary results.

We warmly thank Stefano Capitani for providing numerical results
which enabled us to compare our perturbative results with his and
Yusuke Taniguchi for patiently answering our questions on
perturbative calculations with domain wall fermions.

The development and computer equipment used in this calculation
were funded by the U.S. DOE grant DE-FG02-92ER40699, PPARC JIF
grant PPA/J/S/1998/00756 and by RIKEN. This work was supported by
PPARC grants PPA/G/O/2002/00465, PPA/G/S/\-2002/00467 and
PP/D000211/1. JN acknowledges support from the Japanese Society
for the Promotion of Science.

\bibliographystyle{elsevier}
\vspace{-.5cm}
\bibliography{kaon_da}

\end{document}